\documentclass[twocolumn,amsmath,amssymb]{revtex4}
\usepackage{dcolumn}

\usepackage{hyperref}
\usepackage{graphicx}
\usepackage{dcolumn}
\usepackage{bm}

\usepackage{slashed}

\def\+{\!+\!}
\def\-{\!-\!}
\def\={\!=\!}

\begin{document}

\title{QFT Entanglement Entropy, 2D Fermion and Gauge Fields}

\author{Bom Soo Kim}
\affiliation{%
Department of Mathematics and Physics, University of Wisconsin - Parkside, Kenosha, WI 53144
}%

\date{\today}

\begin{abstract}
\noindent Entanglement and the R\'enyi entropies for Dirac fermions on 2 dimensional torus in the presence of chemical potential, current source, and topological Wilson loop are unified in a single framework by exhausting all the ingredients of the electromagnetic vertex operators of $\mathbb{Z}_n$ orbifold conformal field theory. We employ different normalizations for different topological sectors to organize various topological phase transitions in the context of entanglement entropy. Pictorial representations for the topological transitions are given for the $n=2$ R\'enyi entropy. 

Our analytic computations reveal numerous novelties and provide resolutions for existing issues. We have settled to provide non-singular entanglement entropies that are also continuous across the topological sectors. Surprisingly, in infinite space, these entropies become exact and depend only on the Wilson loop. On a circle, we resolve to find the entropies subtly depend on the chemical potential at zero temperature, which is useful for probing the ground state energy levels of quantum systems. 
\end{abstract}


\maketitle

Entanglement is at the heart of quantum theories, encompassing quantum mechanics, quantum field theories, quantum gravity and quantum information science \cite{Einstein:1935rr}. Entanglement entropy \cite{Horodecki:2009zz,Calabrese:2004eu,Ryu:2006bv,Rangamani:2016dms,QunatumInformation} and its extension, the R\'enyi entropy \cite{Renyi}, can be used to measure quantum information encoded in a quantum state. They were used to probe quantum critical phenomena \cite{Vidal:2002rm}, to classify topological states of matter not distinguishable by symmetries \cite{Kitaev:2005dm}\cite{Levin:2006zz}, and to prove the irreversibility of the renormalization group in 3 dimensional field theories \cite{Casini:2012ei}. Recently, the R\'enyi entropy has been measured in systems of interacting delocalized particles using many-body quantum interference \cite{Experiment}. Despite difficulties in direct computations of entanglement entropy, there have been steady progresses in 1+1 dimensional field theories \cite{Bennett:1995tk,Holzhey:1994we,Calabrese:2009qy,Casini:2009sr}.  

Electromagnetic fields and the corresponding scalar and vector potentials are main tools for manipulating quantum fields. Their time and space components in the 1+1 dimensions are chemical potential $\mu$ and current source $J$, respectively. In the literature, entanglement entropy for free fermions has been claimed to be independent of the chemical potential in finite space at zero temperature \cite{Ogawa:2011bz}\cite{Herzog:2013py} and in infinite space for a single interval \cite{CardySlides:2016}. The R\'enyi entropy in infinite space has been shown to have Wilson loop dependence, yet, unfortunately, has {\it singular} entanglement entropy limit \cite{Belin:2013uta}. (See \cite{Arias:2014ksa} for current source that was considered in a different setup.) These unexpected, and yet fragmented, results deserve systematic investigations with a unified framework that encompasses all these players. Exact and analytic field theoretic results on the entropies can shed valuable insights on the nature of quantum entanglement, but they are scarce. 

\begin{figure}[t]
	\begin{center}
		\includegraphics[width=.43\textwidth, height=0.2 \textheight]{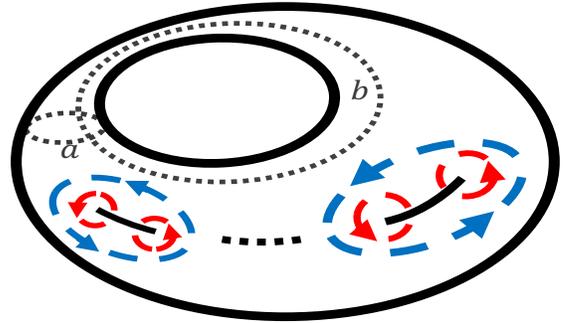} 
		\caption{\small Torus with $n$ replica fermions parametrized by red circles and Wilson loop by blue circles in a single Riemann surface description for R\'enyi entropy. The Wilson loop encircling the cut happens for topologically non-trivial sectors. Accurate pictorial representations are depicted in FIG. \ref{fig:2SheetedContoursCaseA}, \ref{fig:2SheetedContoursCaseB} and \ref{fig:2SheetedContoursCaseC} using $n=2$ sheet example. Chemical potential and current source are parts of the twist boundary conditions of the $a$ and $b$ cycles.   \vspace{-0.25in}
		}
		\label{fig:TorusWithCuts11}
	\end{center}
\end{figure}

In this article, we provide a {\it unified} description of entanglement and the R\'enyi entropies for Dirac fermions on a 2 dimensional torus in the presence of chemical potential $\mu$, current source $J$, and/or topological Wilson loop $w$ based on our previous results \cite{Kim:2017ghc}\cite{Kim:2018xla}. The R\'enyi entropy with $n$-replica fermions perfectly fits into the language of $\mathbb{Z}_n$ orbifold conformal field theories (CFT) and their electromagnetic vertex operators \cite{DiFrancesco:1997nk}\cite{Hori:2003ic}. The Wilson loop parameter $w$ is identified as the CFT's electric parameter (depicted as the blue dashed arrow circle enclosing both ends of a cut in FIG. \ref{fig:TorusWithCuts11}) \cite{Kim:2018xla}, while the replica index $k$, $ k\=(n\-1)/2, (n\-3)/2, \dots,-(n\-1)/2$ for $n$ replica fermions, is identified as the CFT's magnetic parameter (depicted as the red dashed arrow circles around the ends of a cut). In the context of the CFT description, the chemical potential $\mu$ and current source $J$ can be combined with the twist boundary conditions on the $a$, $b$ cycles (also depicted as dotted circles in FIG. \ref{fig:TorusWithCuts11}) on a torus parametrized by $\tau= \tau_1 + i \tau_2\equiv (\alpha + i \beta)/2\pi$. It turns out that the R\'enyi entropy with $\mu, J, w$ along with the replica parameter $k$ exhausts all the ingredients of the $\mathbb{Z}_n$ orbifold CFT, and thus has the most general form for 2 dimensional fermions in the presence of the background gauge fields.  

We perform analytic computations systematically to uncover several novel features and to settle existing issues. Here we focus on two results. First, we resolve to provide non-singular entanglement entropies, that are also continuous across the topological sectors. They overcome the difficulties raised in \cite{Belin:2013uta} and also generalize the results \cite{CardySlides:2016} in the presence of multi-cuts. To provide the possibility of regular entanglement entropy, we use different normalizations for different topological sectors in computing entropies. Surprisingly, in {\it infinite space}, entanglement entropy becomes exact and depends only on the Wilson loop parameter $w$, independent of chemical potential $\mu$ and current source $J$. Along the way, we provide a scheme to systematically characterize various possible topological phase transitions in the context of entanglement entropy with the Wilson loop. 

Second, on a {\it finite circle}, we uncover that entanglement entropy actually depends on the chemical potential $\mu$ at zero temperature in a subtle way, generalizing results of  \cite{Ogawa:2011bz}\cite{Herzog:2013py}\cite{CardySlides:2016}. This happens when the chemical potential is varied and coincides with the ground-state energy spectra of the Dirac fermion, which has integer (periodic) or half integer (anti-periodic) values when the space is compact. Thus entropies can be used to probe the energy spectra of quantum systems. \\

\noindent {\textbf{I. R\'enyi entropy from $\mathbb{Z}_n$ orbifold CFT. }} \\
R\'enyi entropy \cite{Renyi} can be expressed in terms of $n$ replica fermions with appropriate boundary conditions around both ends of finite cuts on a torus \cite{Casini:2005rm}. We find the R\'enyi entropy can be directly constructed by using the electromagnetic twist vertex operators $ \sigma_{w,k}$ of the $\mathbb{Z}_n$ orbifold conformal field theory \cite{DiFrancesco:1997nk} with the following identifications: The electric and magnetic parameters $w$ and $k$ of the vertex operator are associated with the Wilson loop parameter $w$ and the $k$-th replica fermion. The parameter $k$ was previously identified in \cite{Azeyanagi:2007bj}. 

The R\'enyi entropy $S_n$ with $n$ replica fermions is defined as
\begin{align} \label{RenyiEntropy1}
S_n = \frac{1}{1- n} ( \log Z[n] - n \log Z[1] ) \;, 
\end{align}
where the partition function $Z[n]$ is the product of the correlation functions of the vertex operators $\sigma_{w,k} $ \cite{Kim:2018xla} 
\begin{align}
	Z[n]= \prod_{k=-{(n-1)}/{2}}^{{(n-1)}/{2}} \langle \sigma_{w,k} (u) \sigma_{-w,-k} (v) \rangle \;,
\end{align}  
where the magnetic parameter $k= (n-1)/2, (n-3)/2,$ $ \cdots, -(n-3)/2, -(n-1)/2$ is the index representing $n$ replica copies, while the electric one $w$ is the global phase rotation that can be added uniformly to the replica fermions. The points $u$ and $v$ are the end points of the finite cut with length $\ell_t=u\-v$. The normalization factor $ Z[1]$ in \eqref{RenyiEntropy1} plays a central role below. We provide a detailed derivation of the R\'enyi entropy in the presence of the Wilson loop in Supplementary material A. 

As the correlation functions $\langle \sigma_{w,k} (u) \sigma_{-w,-k} (v) \rangle$ factorizes naturally based on the spin structure, the R\'enyi entropy also splits as \cite{Herzog:2013py}\cite{Kim:2017ghc}
\begin{align}
	S^w_{n}=S_{n}^{w,0} + S_{n}^{w,\mu, J} \;. 
\end{align} 
Interestingly, the spin structure independent part $S_{n}^{w,0} $ depends only on the Wilson loop, while the spin structure dependent part $S_{n}^{w,\mu, J}$ depends on all the parameters, $w$, $\mu$ and $J$ along with the fermions' twist boundary conditions on torus. Previously, the magnetic parameter was considered in \cite{Azeyanagi:2007bj}, chemical potential in \cite{Ogawa:2011bz}\cite{Herzog:2013py}, current source in \cite{Kim:2017ghc}, and the Wilson loop in \cite{Belin:2013uta}. All these ingredients are exhausted in formulating our R\'enyi entropy formula as described in the Supplementary material A. As demonstrated below, only $S_{n}^{w,0}$ contributes in infinite space, which can be computed exactly and analytically. 

The conformal dimension of the vertex operator $\sigma_{w,k}$ is given by 
\begin{align} \label{ConformalDimensionBody}
	\Delta_{w,k}=  \frac{1}{2} \alpha_{w,k}^2 \;, \qquad \alpha_{w,k}= \frac{k}{n} + \frac{w}{2\pi} + l^n_{k} \;. 
\end{align} 
The parameter $\alpha_{w,k}$ is bounded as $-1/2 \leq \alpha_{w,k} \leq 1/2$, which was advertised in the context of the R\'enyi entropy in \cite{Belin:2013uta}. A new parameter $l^n_{k}$, depending on both $k$ and $w$, is added to enforce this range of $\alpha_{w,k}$. The finite range of $\alpha_{w,k}$ brings interesting topological transitions \cite{Kim:2018xla} as several different parameters $n, k, w$ participate. Due to this, the Wilson loop parameter $w$ covers the entire real line to accommodate all possible topological winding numbers with $2\pi$ as a unit winding. As $w$ increases to derive $ \alpha_{w,k}$ beyond the bound, $l^n_k$'s develop non-zero integers as the combination of $k$ and $w$ goes beyond the allowed range. This is the basic mechanism of topological transitions in the context of the R\'enyi entropy.  \\ 

\noindent {\textbf{II. Entanglement entropy in infinite space:}} \\
As mentioned above, only the spin structure independent entropy $S_{n}^{w,0}$ survives in infinite space limit. As a function of the ratio $\ell_t/2\pi L$, the sub-system size over the total space, it is given by 
\begin{align} \label{InfiniteREZn}
\log Z^0_w[n]= \big( \sum_{k=-{(n-1)}/{2}}^{{(n-1)}/{2}} \!\! \alpha_{w,k}^2 \big)  \big| \frac{2\pi \eta (\tau)^3 }{\vartheta_1 (\ell_t/2\pi L|\tau)} \big|^{2} \;,
\end{align}
where Jacobi theta and Dedekind functions are $\vartheta _1 (z|\tau) = 2 e^{\pi i \tau/4} \sin (\pi z) \prod_{m=1}^\infty (1- q^m) (1- e^{2\pi i z} q^m )(1 - e^{-2\pi i z} q^m)$ and $\eta(\tau) = q^{{1}/{24}} \prod_{n=1}^\infty (1 - q^n)$, with $ q= e^{2\pi i \tau}$. This result is involved with the generalized conformal dimension Eq. \eqref{ConformalDimensionBody} \cite{Kim:2018xla}. For readers' convenience, the derivation is presented in the Supplementary material A. 

Novelty for the R\'enyi entropy in the presence of the Wilson loop comes from the topological transitions depending on $k$ and $w$. As $w$ increases and derives $\alpha_{w,k}$ beyond the allowed range $| \alpha_{w,k}| < 1/2$, some of the $l_k$'s are adjusted to be non-zero to keep the parameter in the range. For $n=2$, there are two replica copies, $k=\pm 1/2$. When $w > \pi/2$, $\alpha_{w,1/2}  > 1/2+ l_{1/2}$ for $k=1/2$. Thus Replica fermion with $k\= 1/2$ develops a phase shift, $l_{1/2}=-1$. For $\pi/2 < w < 3\pi/2$, both $l_k$'s become non-zero, $l_{1/2}=l_{-1/2} =-1$. And the transitions go on.

In general, for a given $n$, topological transitions happen when $w= (2Q+1\!) \pi/n$ with an integer parameter $Q= 0,1,2,\cdots$. The corresponding topological level has a range $(2Q-1 ) \pi/n  \leq w < (2Q+1) \pi/n$, in which the phase factors $l^n_{k}$ are 
\begin{align}
	l^n_{k} = - \text{Floor} [1/2+(Q+k)/n] \;,  
\end{align}
where $\text{Floor}[~]$ rounds down to the nearest integer. Note, for the first $n$ transitions, the phases becomes non-zero, $l_k= -1$, starting from the maximum $k=(n\-1)/2$ to the minimum $k= (1-n)/2$, followed by another $n$ transitions with $l_k= -2$ with the same order, as illustrated in the table \ref{Table1}.

\begin{table}[!t] 
	\centering
	\begin{tabular}{| c || c | c || c | c | c || c | c | c | c || c | c | c | c | c || c } 
		\hline 
		$Q$  & $l^2_{1/2}$ & $l^2_{-1/2}$ & $^3l_{1}$ & $l^3_{0}$ & $l^3_{-1}$ & $l^4_{3/2}$ & $l^4_{1/2}$ & $l^4_{-1/2}$ & $l^4_{\!-3/2}$  & $\cdots$ \\ [0.5ex] 
		\hline\hline 
		0 & 0 & 0 & 0 & 0 & 0 & 0 & 0  & 0 & 0 & \!$\cdots$\!  \\ 
		1 & -1 & 0 & -1 & 0 & 0 & -1 & 0 & 0 & 0  & \!$\cdots$\!  \\ 
		2 & -1 & -1 & -1 & -1 & 0 & -1 & -1 & 0 & 0 & \!$\cdots$\!  \\ 
		3 & -2 & -1 & -1 & -1 & -1  & -1 & -1 & -1 & 0 & \!$\cdots$\!  \\ 
		4 & -2 & -2 & -2 & -1 & -1  & -1 & -1 & -1 & -1 & \!$\cdots$\!  \\ 
		$\vdots$ & $\vdots$ & $\vdots$ & $\vdots$ & $\vdots$ & $\vdots$ & $\vdots$ & $\vdots$ & $\vdots$ & $\vdots$ & $\vdots$\\
		\hline
	\end{tabular}
	\caption{Specific values of the phase factors $l^n_{k}$ for a given parameter $Q$ that describes the topological transitions.   
	}
	\label{Table1}
\end{table}

Non-vanishing $l_k$'s put challenges both on computing the sum in \eqref{InfiniteREZn} and on determining the normalization factor $Z[1]$ in $ S_{n}^{w,0}$. Moreover, {\it a priori}, it is not obvious how to choose the normalizations $Z[1]$ for different topological sectors. The QCD $\theta$-vacua is a canonical example that uses different normalizations for different topological sectors when computing correlation functions \cite{Callen1976}.  

We realize $Z[1]$ can be used to systematically investigate the properties of the topological transitions in the context of the entropies.  
\begin{align}\label{NormalizationFactorG}
	\log Z[1] = \lim_{n\to 1} \left( \log Z[n] \right) + \sum_{q=1}^{\infty}  (n-1)^q \log \alpha^w_q
	\;.
\end{align}
The first part guarantees the R\'enyi entropy to have a smooth entanglement entropy limit. With this form, the normalization factor depends on both $Q$ and $w$. 

The second part of \eqref{NormalizationFactorG} can model the ``order'' of phase transitions across different topological sectors \cite{Charmousis:2010zz}%
\footnote{One can justify the possible existence of this extra contribution because the partition function and the R\'enyi entropy are discontinuous at transition points, which are not directly associated with the parameter $n$ and the limit $n \to 1$. Let us illustrate this with a function $f(x,y)$. 
We can expand the function around $x=1$ as $ f(x, y) = f(1,y) + (x-1) f'(1,y) + (x-1)^2 f''(1,y)/2 + \cdots $ with $'$ being $\frac{\partial}{\partial x }$. 
If $ f(x=1,y)$ is smooth as a function of $y$, then the limit $f(1,y)$ would be clear. 
Yet, if the function $ f(x, y) $ is discontinuous at $y=y_0$, it is not clear which values we need to choose at that particular point. This property is due to the discontinuity of the 
function $ f(x, y) $ in the parameter $y$, not $x$. }. 
For example, the entropies can be continuous, signaling a first order transition, by choosing $\alpha^w_{q=1}$ (denoted as $\tilde \alpha_w$ below) appropriately. We can achieve the second order transition using $\alpha^w_{q=2}$, so that the first derivatives become also continuous. Higher order transitions can be achieved by including more terms in the expansion \eqref{NormalizationFactorG}. 
From theoretical standpoint, all different topological transitions are probably, and thus there is no preferred way to fix these coefficients for a given system without experimental inputs. 

Previously, $\log Z[1]$ was chosen uniformly across different topological sectors in \cite{Belin:2013uta}, and the R\'enyi entropy did not provide entanglement entropy as its $n\to 1$ limit is singular. On the other hand, if one chooses $\log Z[1] = \lim_{n\to 1} \left( \log Z[n] \right)$, the R\'enyi entropy provides the corresponding entanglement entropy, which is discontinuous at the topological transition points \cite{Kim2017Unpublished}. 

Here we choose {\it the first order phase transition scheme}, 
\begin{align} \label{NFFirstOrder}
\log Z[1] = \lim_{n\to 1} \left( \log Z[n] \right) + (n-1)\log \tilde \alpha_w \;,
\end{align}
along with the conditions:  
\begin{itemize}
	\item[I.] The R\'enyi entropy has a smooth entanglement entropy limit for $n \to 1$.
	\item[II.] The R\'enyi and entanglement entropies are continuous across different topological sectors.
\end{itemize}
Note $ \tilde \alpha_w$ ($\alpha_{q=1}^w$ in \eqref{NormalizationFactorG}) is determined by the condition II. $ \tilde \alpha_w$ ensures continuity of entropies at the topological transition points by adapting different normalizations for different topological sectors as explained above. Detailed computations can be found in \cite{Kim:2018xla}. 

\begin{figure}[b]
	\begin{center}
		\includegraphics[width=.45\textwidth]{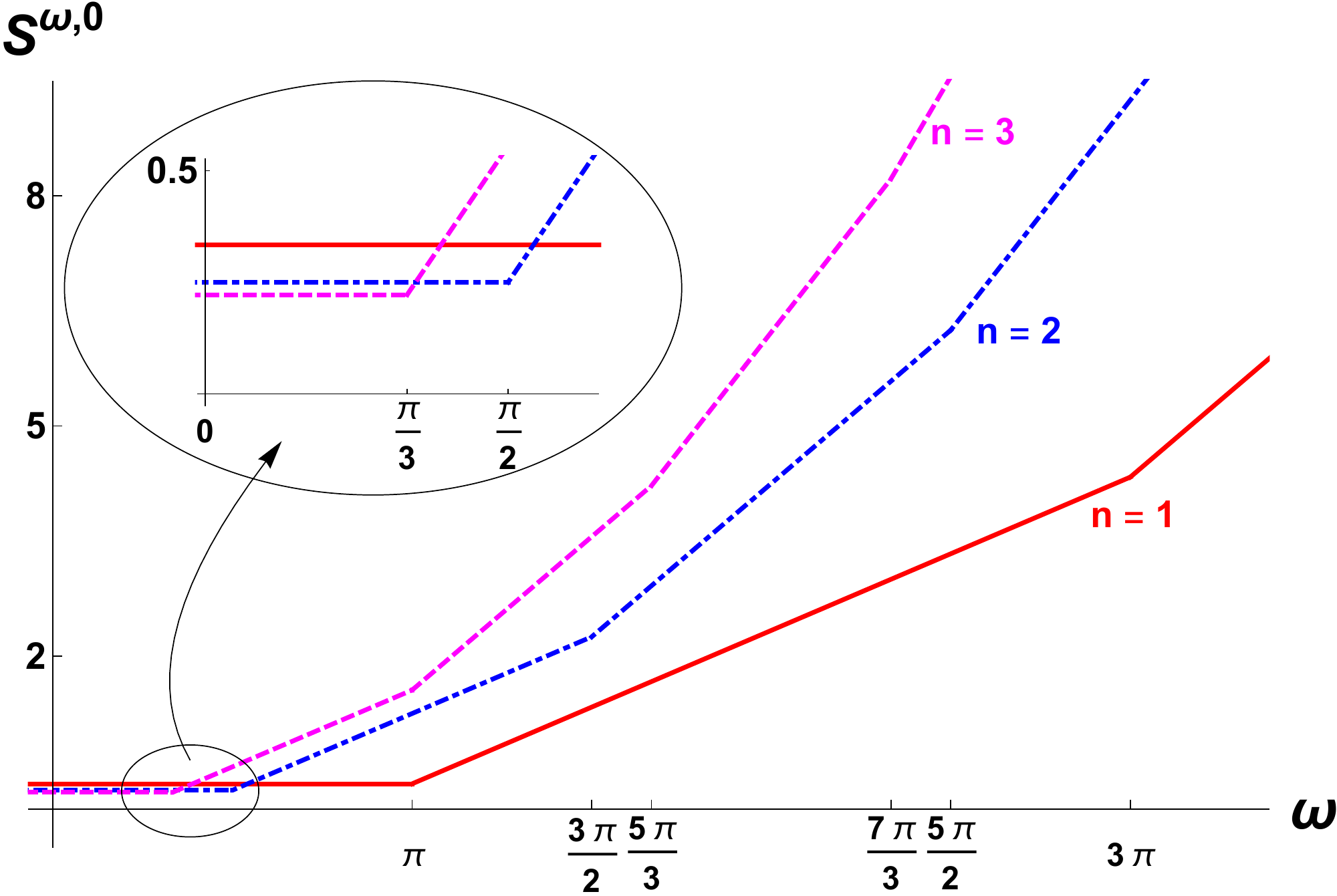}
		\caption{\footnotesize\small R\'enyi entropies given in \eqref{Cnwp}, $S_{n}^{w,0}$, for $n=3,2,1$ as a function of $w$ across the first three topological sectors. The first order transitions are clearly visible. Here we only present the positive values of $w$.   \vspace{-0.25in}
		}
		\label{fig:RenyiWilsonloop}
	\end{center}
\end{figure}

\begin{figure}[t]
	\begin{center}
		\includegraphics[width=.46\textwidth]{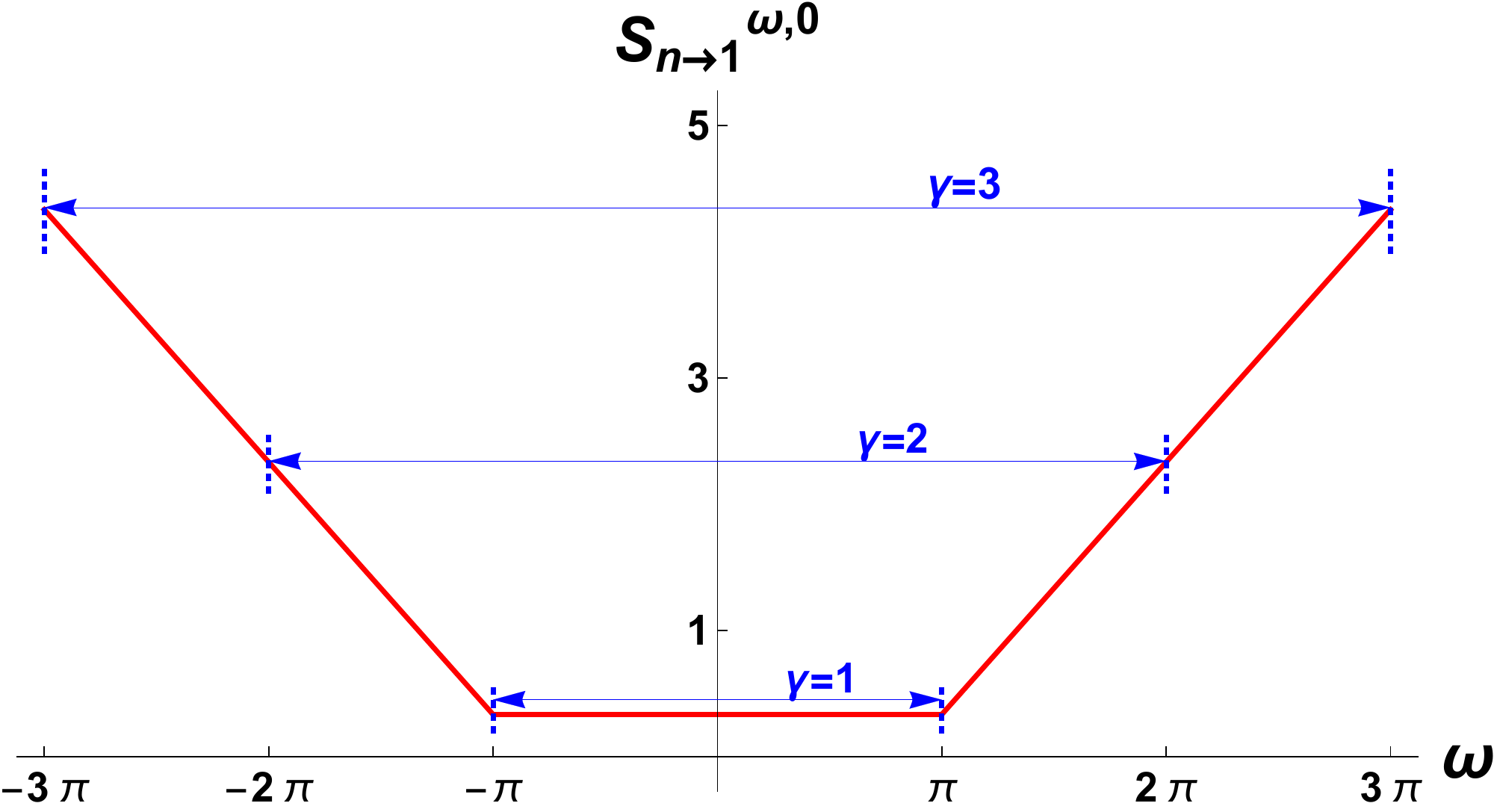} 
		\caption{\footnotesize\small Entanglement entropy given in \eqref{InfiniteEE}, $S_{n=1}^{w,0}$, as a function of $w$ across the first three topological sectors $\gamma=1,2,3$ with  corresponding domains $-\gamma \pi \leq w < \gamma \pi $.     \vspace{-0.25in}
		}
		\label{fig:EntanglementWilsonloop}
	\end{center}
\end{figure}

By adapting this scheme, we compute the R\'enyi entropy iteratively using \eqref{InfiniteREZn} and \eqref{NFFirstOrder} \cite{Kim:2018xla}. 
In infinite space ($L \!\to\! \infty $), 
\begin{align} \label{Cnwp}	
S_{n,Q}^{w,0} =2 \bigg( \frac{n+1}{12 n} + \frac{Q w}{\pi} - \frac{Q^2}{n} \bigg) \log |\ell_t|  \;.
\end{align} 
The result is valid for ${(2Q- 1)\pi}/{n} \leq w < {(2Q + 1)\pi}/{n} $, the range of the topological sector parameterized by $Q$. Three cases $n=3,2,1$ are depicted in the FIG. \ref{fig:RenyiWilsonloop}. One can clearly see that the topological transitions are more frequent for larger $n$ with more replica fermions. 

Entanglement entropy can be obtained by the smooth $n\! \to\! 1$ limit, overcoming the difficulties reported in \cite{Belin:2013uta}. 
\begin{align} \label{InfiniteEE}
S_{n=1,Q}^{w,0} \= 2 \bigg( \frac{Qw}{\pi} - \frac{6Q^2 -1}{6} \bigg) \log |u-v|   \;,
\end{align}
which is valid for $ (2Q\- 1)\pi  \leq  w < (2Q\+ 1)\pi $. Entanglement entropy is continuous and positive definite. See the FIG.  \ref{fig:EntanglementWilsonloop}. For the fundamental domain $ -\pi \leq w<\pi$ of the Wilson loop ($\gamma=1$), entanglement entropy is only defined on the domain and independent of $w$. By including larger range of $w$, we can see the non-trivial entanglement profile that depends on $w$.  

The entropies \eqref{Cnwp}\eqref{InfiniteEE} are exact in infinite space because the rest of the entropy (spin structure dependent part) vanishes as fast as $\mathcal O(\ell_t^2/L^2)$ as $L\!\to\! \infty$ \cite{Kim:2017ghc}\cite{Kim:2018xla}.  This is one of our main results, achieved by adapting proper normalization factor \eqref{NormalizationFactorG} that accommodates the first order phase transitions. \\

\begin{figure}[b]
	\begin{center}
		\includegraphics[width=.235\textwidth]{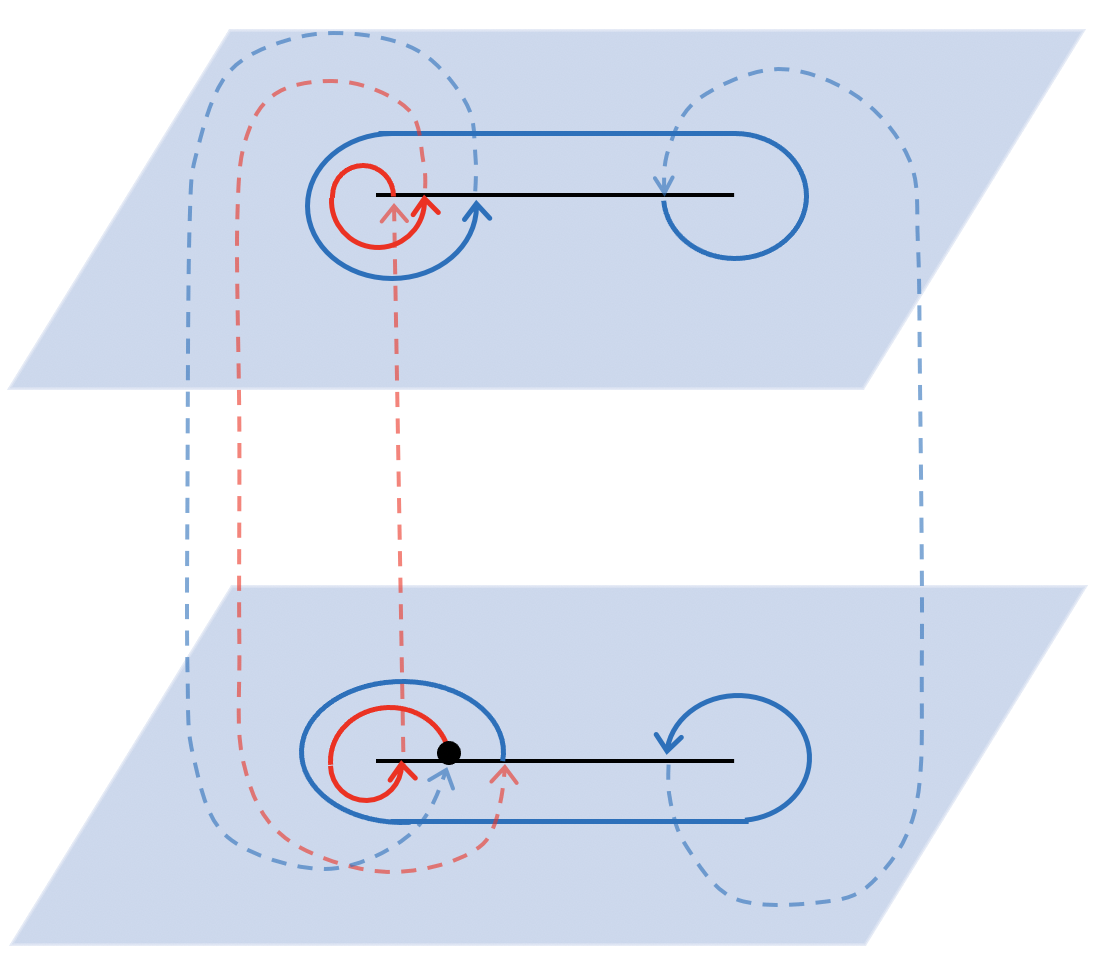} 
		\includegraphics[width=.235\textwidth]{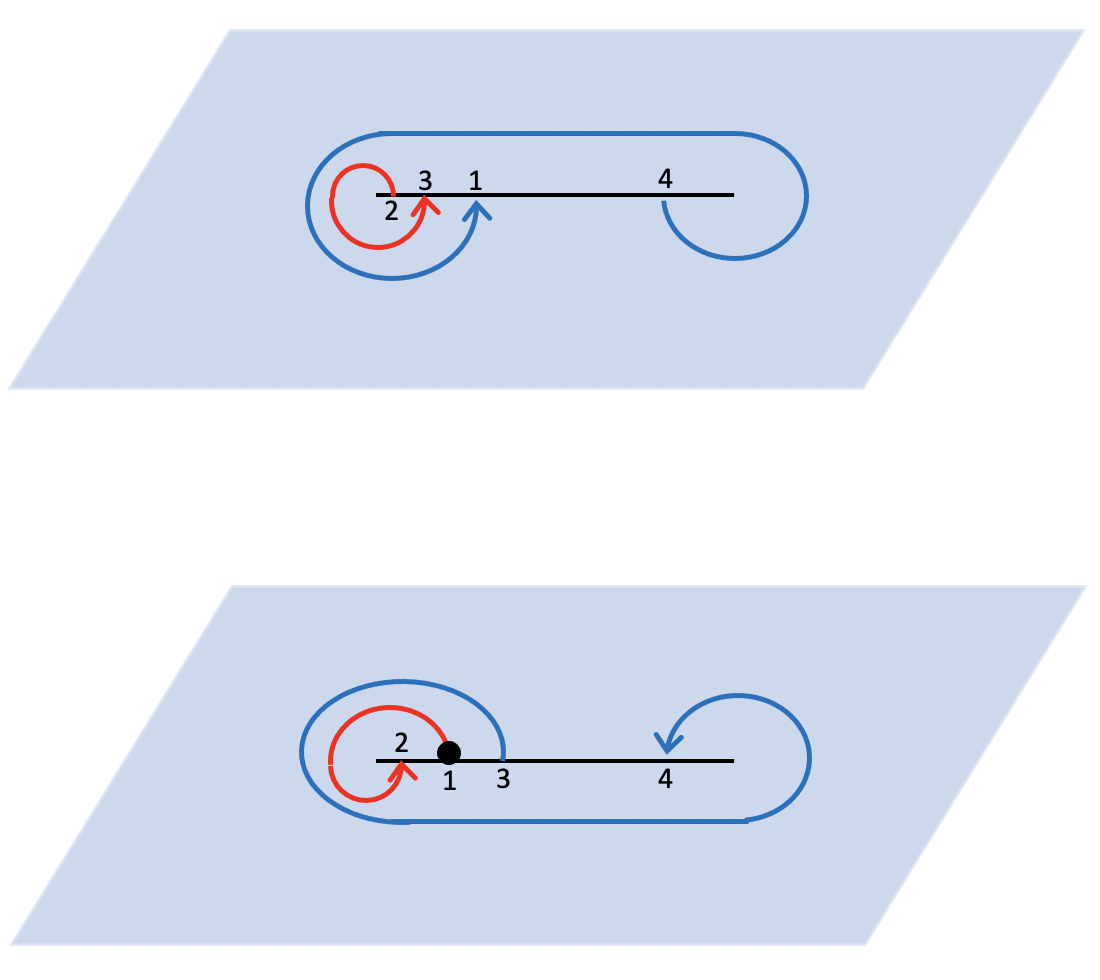} 
		\caption{\footnotesize\small $n=2$ Riemann sheets with the Wilson loop circling around once for both cuts. The contour circles around their left ends counter-clockwise.   \vspace{-0.25in}
		}
		\label{fig:2SheetedContoursCaseA}
	\end{center}
\end{figure}

\begin{figure}[t]
	\begin{center}
		\includegraphics[width=.235\textwidth]{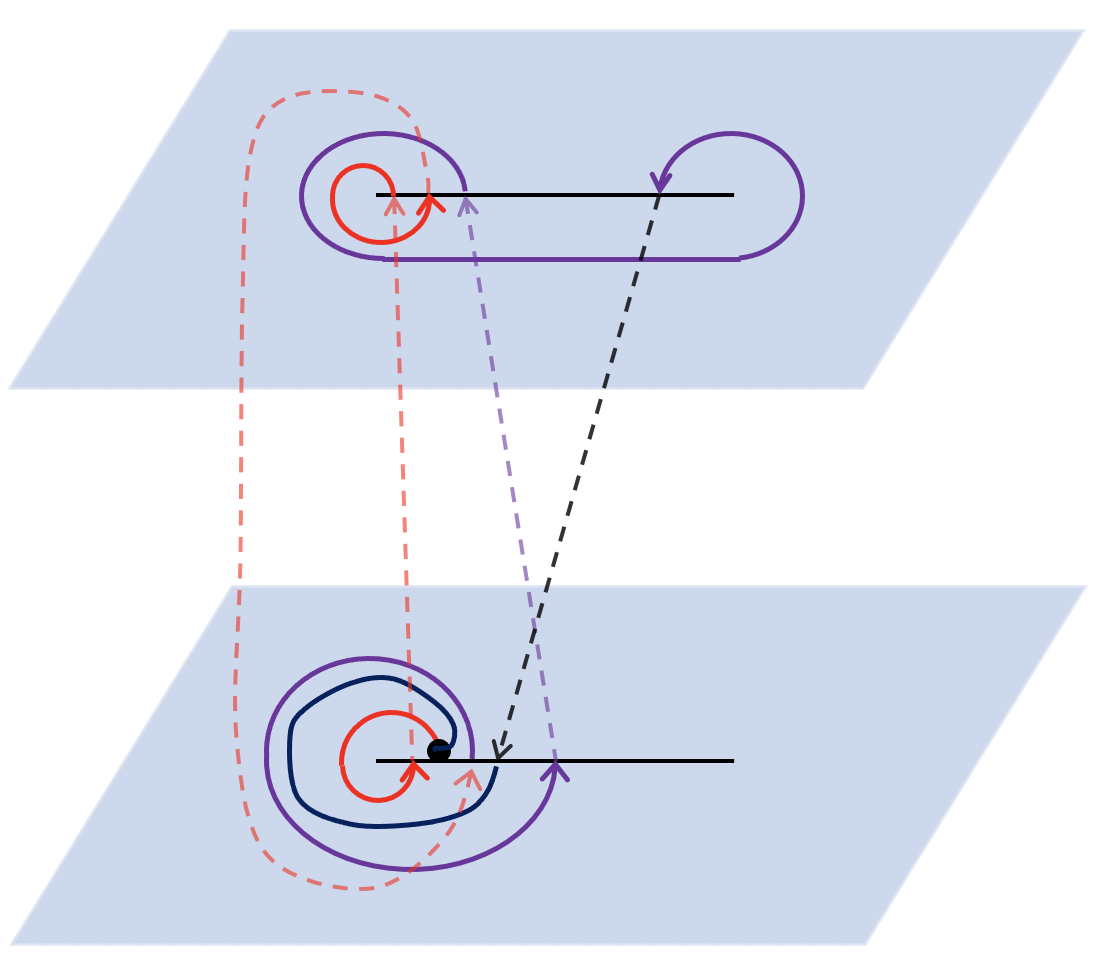} 
		\includegraphics[width=.235\textwidth]{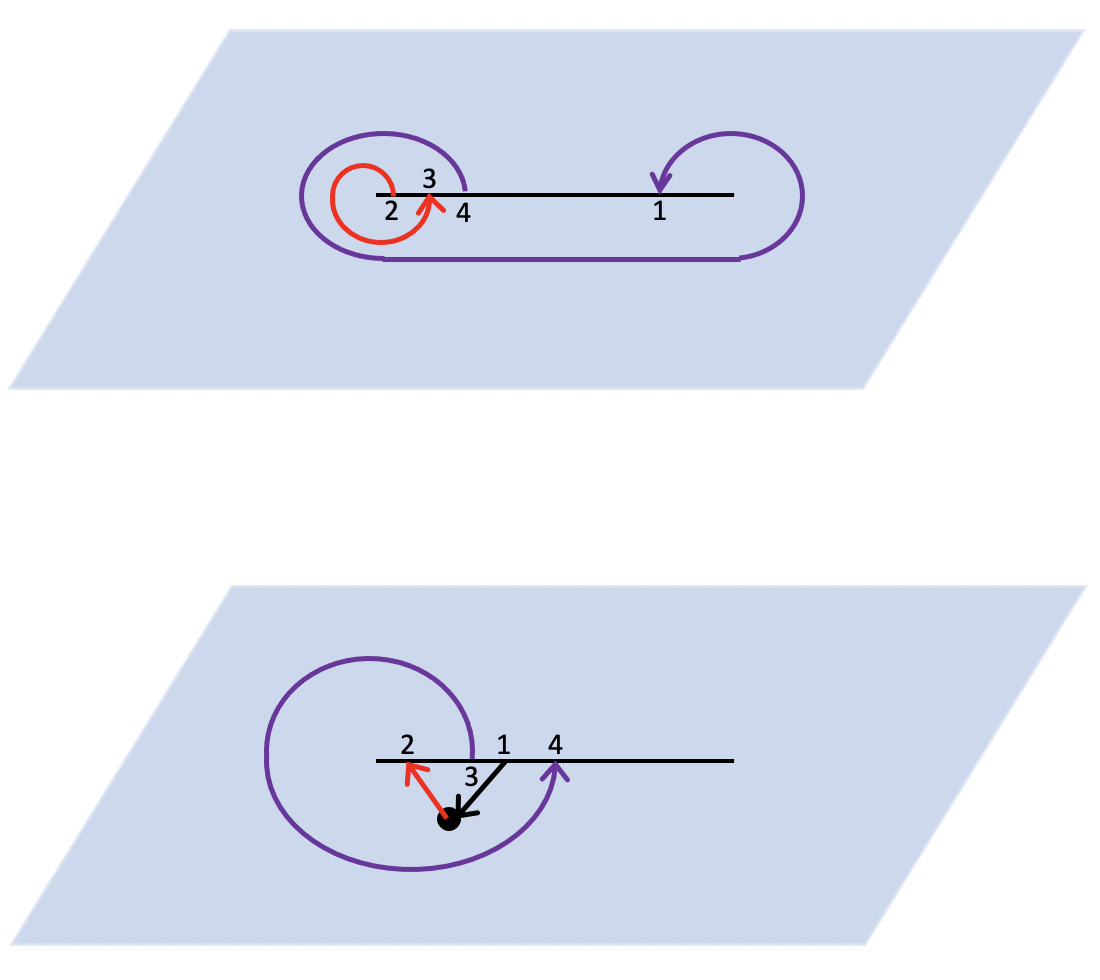} 
		\caption{\footnotesize\small $n=2$ Riemann sheets with the Wilson loop circling around once for only the upper cut. In the lower cut, the red and black circles combines to reduce to be a topologically trivial point. Switching the upper and the lower Riemann sheets describes the topologically equivalent R\'enyi entropy.    \vspace{-0.25in}
		}
		\label{fig:2SheetedContoursCaseB}
	\end{center}
\end{figure}

\begin{figure}[b]
	\begin{center}
		\includegraphics[width=.235\textwidth]{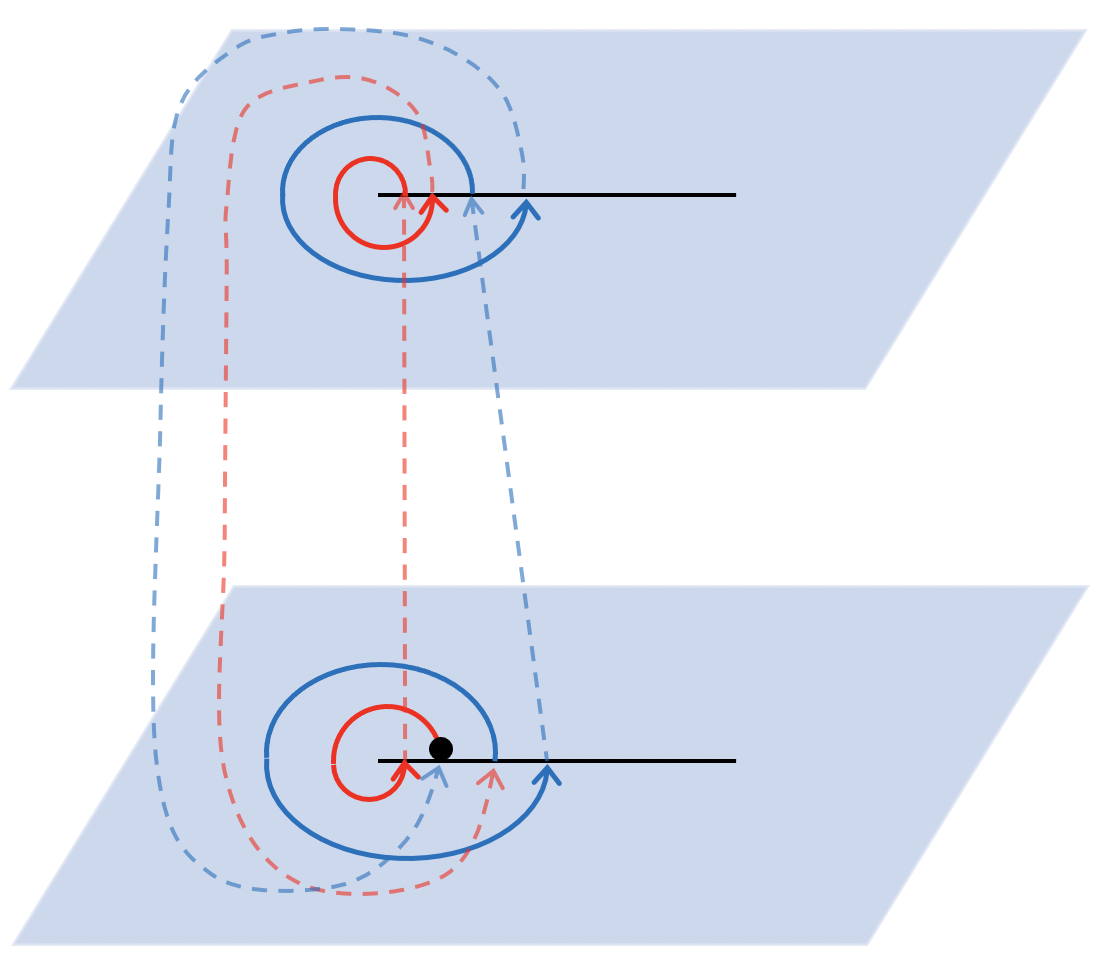} 
		\includegraphics[width=.235\textwidth]{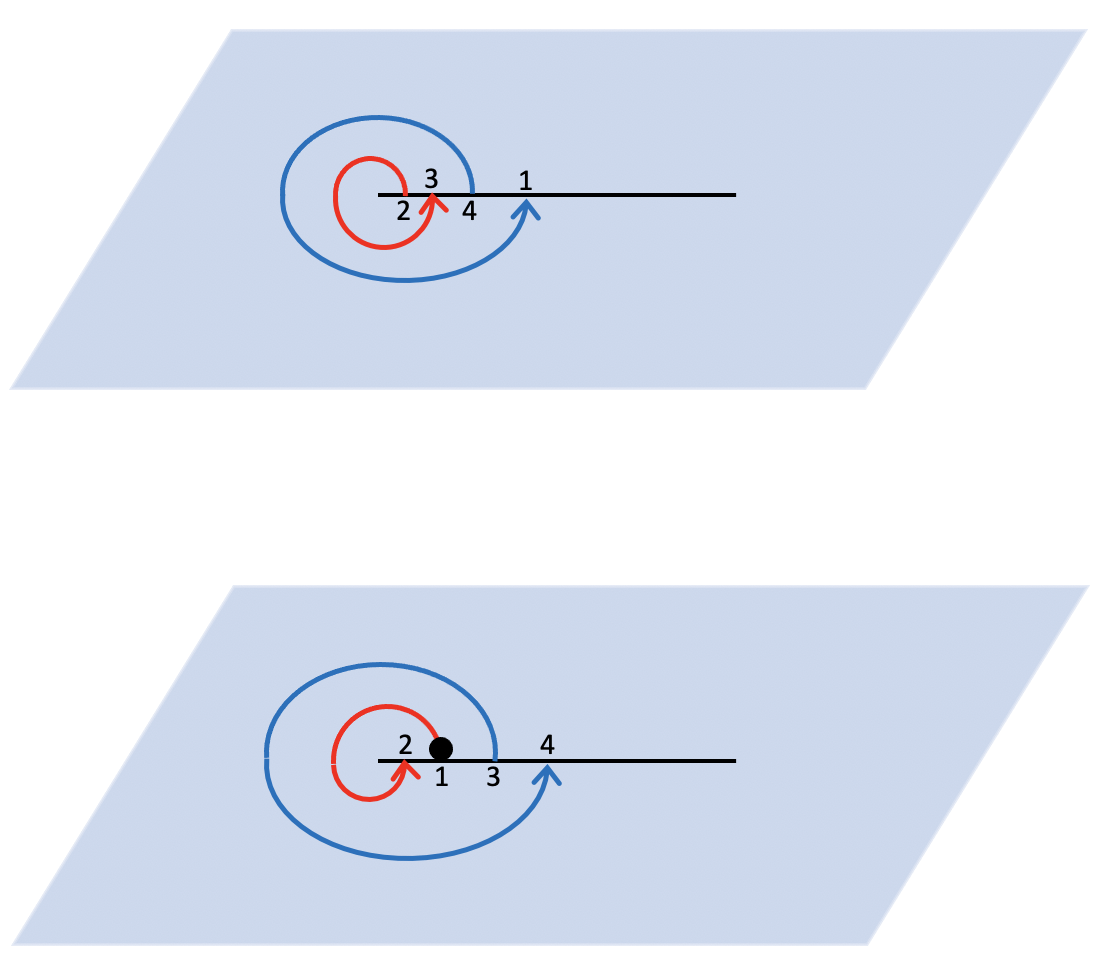} 
		\caption{\footnotesize\small $n=2$ Riemann sheets for topologically trivial sectors after taking of the reduction with Wilson loop. The range of the Wilson loop parameter is reduced, which takes care of the range of one topological sector.   \vspace{-0.25in}
		}
		\label{fig:2SheetedContoursCaseC}
	\end{center}
\end{figure}

\noindent  {\it $A.$ Example: $n=2$ sheeted Riemann surfaces.} \\ 
Here we visualize different topological sectors for $n=2$ and $ -\pi \leq w < \pi$ ($\gamma=1$). This is the case when the blue dashed line circles around the cut {\it once} in FIG. \ref{fig:TorusWithCuts11}. As mentioned above, there are two transition points located at $w = \pm \pi/2 $ and thus three distinct topological sectors. 

Here we adapt the $n=2$ sheeted Riemann surface instead a single Riemann surface and use a different periodic range $ 0 \leq w < 2\pi$ for clear understanding of topological transitions (while this change obscures its connection to entropy formulas). We focus on the connected path circling around the left end of the cut with a counter-clockwise direction. Starting from the black dot, one can construct the contour as in left panel of FIG. \ref{fig:2SheetedContoursCaseA}. The red circles are the same contours without the Wilson loop contribution. The right panel depicts the same connected paths replacing the dashed lines with the same numbers on the two Riemann sheets. The blue contour is topologically equivalent to circling the entire cut counter-clockwise once. FIG. \ref{fig:2SheetedContoursCaseA} is the only topologically inequivalent contour that encircles the cuts of both Riemann surfaces.  

We can systematically reduce the number of Wilson loop circling around the cuts by choosing the purple contour to only encircle the left end of the cut in the lower Riemann sheet as in FIG. \ref{fig:2SheetedContoursCaseB}. 
There is a topologically equivalent configuration, switching the upper and lower Riemann sheets. One can take a linear combination of these two topologically equivalent configurations to establish symmetric configuration for upper and lower Riemann sheets. Note the contour described in FIG. \ref{fig:2SheetedContoursCaseB} is topologically inequivalent compared with that in \ref{fig:2SheetedContoursCaseA} as we cannot continuously deform the contour of FIG. \ref{fig:2SheetedContoursCaseA} to the one of FIG. \ref{fig:2SheetedContoursCaseB}, describing the topological transition in a pictorial way. 

FIG. \ref{fig:2SheetedContoursCaseC} depicts the case when the contours around the upper and the lower cuts are topologically trivial. This can be achieved by encircling the Wilson loop only the left ends of cuts. By doing so, the range of the topological Wilson loop parameter is reduced. This is possible as the topological transitions also involve with the other parameter $k$. For example, $w=\pi/2$ along with $k=1/2$ achieve to saturate $\alpha_{w,k} =1/2 $, a transition point. 

We emphasize that our visualizations are done by adapting the range $0 \leq w < 2\pi $. It has advantages in depicting the contours as we can consistently use the counter-clockwise direction when circling around the cuts. But the connections to the explicit computations and the transition points located at $w= \pm \pi/2$ for the range $-\pi \leq w <\pi $ are obscure. The three different topological sectors for $n=2$ separated by $ w = \pm \pi/2$ are related to these three figures. FIG.  \ref{fig:2SheetedContoursCaseC} and FIG. \ref{fig:2SheetedContoursCaseA} are connected to the topological sectors $-\pi \leq w < -\pi/2 $ and $\pi/2 \leq w <\pi $, respectively. And FIG. \ref{fig:2SheetedContoursCaseB} is related to the sector $-\pi/2 \leq w < \pi/2 $. The middle sector seems to be twice bigger than the other two. It is clear that for $\gamma=2$ case, the middle sectors have twice larger ranges than the end sectors. We mention that these visualizations for topological loop contours are valid for infinite space and also for a finite circle. \\

\noindent {\textbf{III. Entanglement entropy on a circle:}} \\
When replica fermions are on a finite-size torus, the spin structure dependent parts of the entropies (SSDE) do depend on all the players including $\mu$, $J$, and $w$, along with the two, $a$ and $b$, cycles of torus. 

Fermion described by a wave function $\psi$ has twist boundary conditions (BC) on torus with coordinates $ (s + it )/2\pi$ and moduli $\tau = \tau_1 + i \tau_2$, 
\begin{align} \label{TwistedeBCBody}
\begin{split}	
	\psi (t,s) &= e^{-2\pi i \tilde a} \psi (t,s+2\pi) \\
	&= e^{-2\pi i \tilde b} \psi (t+2\pi \tau_2, s+2\pi \tau_1)  \;,
\end{split}	
\end{align}
where $\tilde a, \tilde b$ are twist parameters along the cycles. The fermions with twist BC has an equivalent description as the fermions with trivial BC under the influence of background gauge field $\mu$ and $J$ \cite{Hori:2003ic}\cite{Kim:2017ghc}. The equivalence is 
\begin{align}
	A = i\mu~ dt + J~ ds = \frac{\tilde b - \tilde a\tau_1}{\tau_2} ~ dt + \tilde a~ ds \;. 
\end{align}
Thus, chemical potential $\mu$ and current source $J$ can be integrated to the partition function of the replica fermions by modifying $\tilde a \to \tilde a+ J $ and $\tilde b \to \tilde b + \tau_1 J + i\tau_2 \mu $.

This observation enables us to construct the general formula for the R\'enyi entropy. For simplicity, we focus on the anti-periodic BC for both spatial and temporal circles (known as the NS-NS sector), which is to set $\tilde a+J=1/2, \tilde b=1/2$. The SSDE $S_{n}^{w,\mu, J} $ can be obtained by 
\begin{align} \label{FiniteREZn}
\log Z^{\mu,J}_w = \!\! \sum_{k=-\frac{n-1}{2}}^{\frac{n-1}{2}} \!\! \log \big| \frac{\vartheta_3 (\alpha_{w,k} \frac{\ell_t}{L} \+ \tau_1 J \+ i \tau_2 \mu|\tau)}{\vartheta_3 (\tau_1 J + i \tau_2 \mu|\tau)} \big|^2 \;.
\end{align} 
Here 
$
\vartheta_3 (z|\tau) = \prod_{m=1}^\infty (1 - q^m)(1 + y q^{m-\frac{1}{2}})(1 + y^{-1} q^{m-\frac{1}{2}})$, where $q\=e^{2\pi i \tau}$ and $y\=e^{2\pi i z}
$. 
Periodic fermions with $\tilde a +J=0$ show similar properties. This can be derived using the formulas in the Supplementary material A.  Hereafter, we focus on the low temperature limit, $\beta\to \infty$. 
\\

\noindent  {\it $A.$ Chemical potential dependence ($\mu$):} \\
To study $\mu$ dependence of the entropies, we set $w= 0$ along with $\tau_1 =J =0$. The normalization factor becomes trivial, $Z[1]=1 $ for this trivial sector. Then the R\'enyi entropy \eqref{FiniteREZn} reduces to 
\begin{align} \label{SAmu3}
S^\mu_n &= \frac{1}{1\-n} \Big[ \sum_{k=-\frac{n-1}{2}}^{\frac{n-1}{2}} \!\!\! \log \big|\frac{\vartheta_3 (\frac{k}{n} \frac{\ell_t}{2\pi L}\+\frac{i\beta\mu}{2\pi} |\frac{i\beta}{2\pi} )}{\vartheta_3 (\frac{i\beta\mu}{2\pi} |\frac{i\beta}{2\pi})} \big|^2 ~\Big] \;. 
\end{align}  
In the low temperature limit, we use $\log (1 + y q^{m-1/2}) = \sum_l (-1)^{l-1} y^l q^{l(m-1/2)}/l$ (for $|q|= e^{-\beta} \ll 1$) to expand $\vartheta_3$. Combining terms with the same $l$, followed by sum over $k$ and $m$, we get entanglement entropy $(n=1)$
\begin{align} \label{SAmuNS}
	S^\mu_{1} 
	&=2 \sum_{l=1}^{\infty} \frac{(-1)^{l-1}}{l} 
	\frac{\cosh (l\beta\mu)}{\sinh (l\beta/2 )} \big[1 - \frac{l \ell_t}{2L} \cot \big(\frac{l \ell_t}{2L} \big) \big] \;.
\end{align} 
Detailed computation can be found in Supplementary material B. 
The result is valid for $e^{-\beta\mu - \beta/2} < 1$ and $e^{\beta\mu - \beta/2} < 1 $, and thus for $-1/2 < \mu < 1/2$. $\mu$ dependent entropies seem to vanish both at zero temperature $\beta \to \infty$ and in infinite space $L\to \infty$, which were argued previously \cite{Ogawa:2011bz}\cite{Herzog:2013py}. Yet, the argument is not valid for $\mu = \pm 1/2$.  

We consider two limits $\beta \to \infty$ and $\mu \to 1/2$ together while maintaining their product $\beta(\mu-1/2) = M$ held fixed. A modified expansion for $\vartheta_3$ is useful. $\vartheta_3 (z|\tau) = (1 + y^{-1} q^{\frac{1}{2}}) \prod_{m=1}^\infty (1 - q^m)(1 + y q^{m-\frac{1}{2}})(1 + y^{-1} q^{m+\frac{1}{2}}) $. The front factor yields a finite contribution in the limits as $(1 + y^{-1} q^{\frac{1}{2}}) = 1+ e^M$. While the entropies exist for general $M$, we consider $M< 1$ for simplicity. 
Then, the middle factor in \eqref{SAmuNS} turns into
\begin{align} \label{FiniteTerm21}
\frac{\cosh (l\beta\mu)}{\sinh (l\beta/2 )} ~~ \rightarrow ~~ e^{-Ml} + \frac{e^{-l\beta/2}}{\sinh(l\beta/2)} \;. 
\end{align} 
To get this result, we use $|y q^{1/2}| < 1$ and $|y^{-1} q^{3/2}| < 1$, that is valid for a different range of $\mu$, $-1/2 < \mu < 3/2$. The term $e^{\-Ml}$ is finite and survives at zero temperature. 

This phenomenon is more general. There are non-vanishing and finite contributions whenever the product $\beta \left(\mu - (2N+1)/2 \right) $ is held fixed for the limits $\beta \to \infty$ and $\mu \to (2N+1)/2$ for an integer $N$. This parameter $N$ is identified as the energy levels of the fermions in a compact circle with an anti-periodic BC. This story is also true for the periodic fermions ($\tilde a=0$). Their entropies pick up finite values when we take the limits $\mu \to N$ and $\beta \to \infty$ simultaneously. Combining the periodic and anti-periodic fermions, we conclude the entropies have non-trivial chemical potential dependences when  
\begin{align}
\beta \Big(\mu - \frac{N}{2} \Big) = const. \;, 
\end{align}
for $\beta \to \infty$ and $\mu \to N/2$. This can be used to probe the ground state energy levels of quantum systems by varying the chemical potential. We expect this happens generically. It is our second main result. 

We further take the large radius limit, $L \to \infty$. We sum over $l$ in \eqref{SAmuNS} with the result \eqref{FiniteTerm21}. Then, 
\begin{align}
S^M_{1} = \frac{(\ell_t/2L)^2}{3 [1+ \cosh (M)]} + \mathcal O \Big(\frac{\ell_t}{2L} \Big)^4 \;.
\end{align}
We see that the chemical potential dependence of entropies vanish in infinite limit, which is mentioned above. Furthermore, the current source $(J)$ dependence of entropies also vanishes in infinite space limit. \\  

\begin{figure}[t]
	\begin{center}
		\includegraphics[width=.45\textwidth]{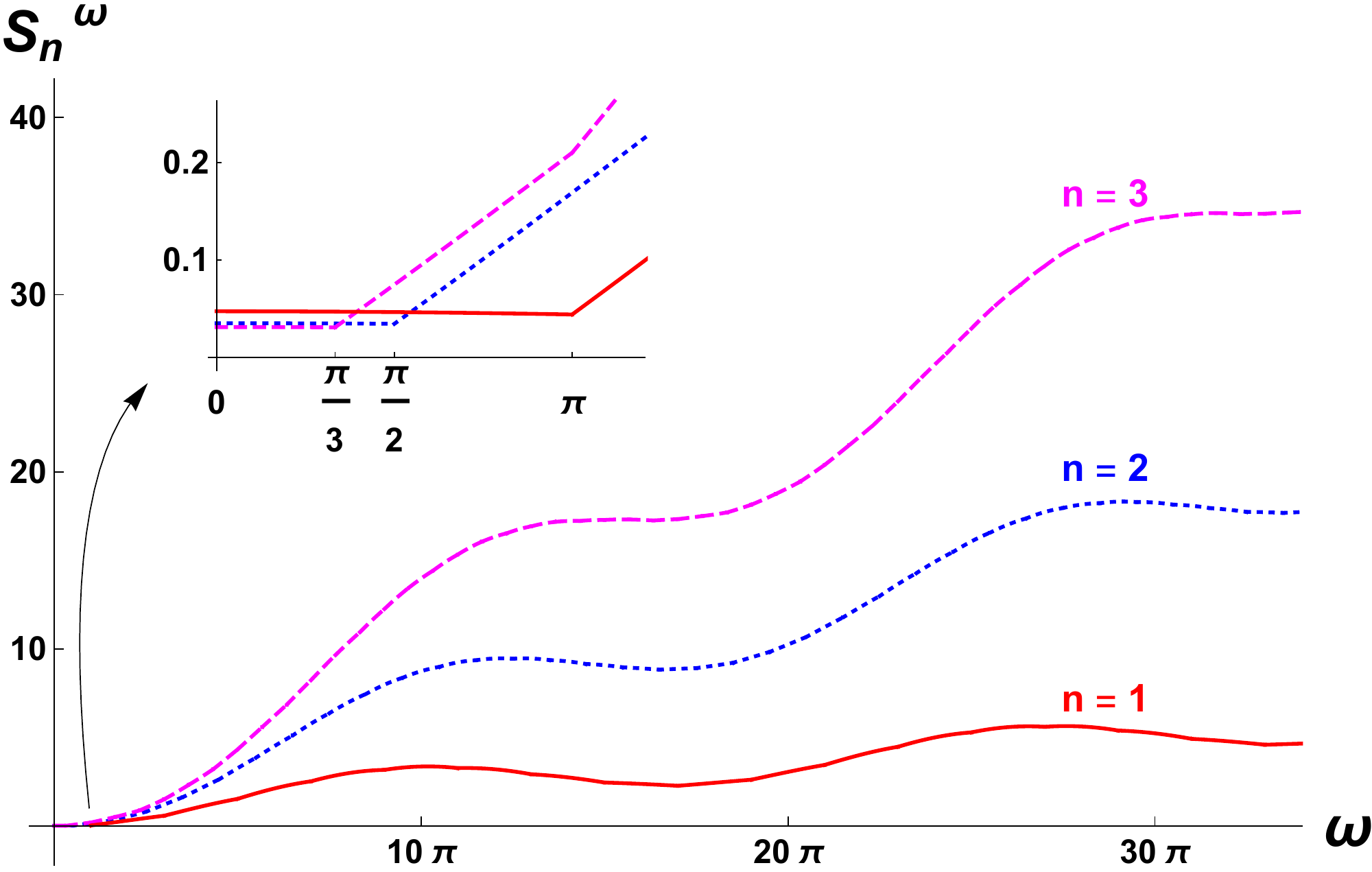} 
		\caption{\footnotesize\small The Renyi entropy $S_{n}^{w,\mu=0,J=0}$ for $n=3,2,1$ ($n=1$ given in \eqref{LowTEEW}) as a function of Wilson loop parameter $w$. The inset is an enlarge part of the fundamental domain of $w$ that reveals the topological transitions. \vspace{-0.25in}
		}
		\label{fig:RenyiLowT2}
	\end{center}
\end{figure}

\noindent {\it $B.$ Wilson loop dependence ($w$):} \\
To see the Wilson loop $w$ effects clearly in SSDE, we choose $\tau_1 \=\mu \= J \=  0$. Then, \eqref{FiniteREZn} and \eqref{NFFirstOrder} (for the {\it first order topological transition}) are simple to evaluate following the previous steps. At zero temperature, entanglement entropy has the form. 
\begin{align} \label{LowTEEW}
\begin{split}
&S_{1} =\sum_{l=1}^{\infty}  \frac{2 (-1)^{l-1}}{l \sinh (\frac{l\beta }{2} ) } 
\Big[ \cos ( \frac{w R_l}{\pi } )  \big[ 1 - R_l \cot ( R_l ) \big] \\ &\quad + 2Q R_l \sin ( R_l )  
 + 2\sin ( Q R_l )  \sin ( [ \frac{ w}{\pi  } \- Q] R_l ) \\
 &\quad \- 2  R_l \sin ( Q R_l )  \sin ( \frac{ w \- \pi Q}{\pi}  R_l )  \cot ([w \- Q\pi] R_l )  \Big], 
\end{split}
\end{align} 
where $R_l = l \ell_t /2L$. The result is valid for $(2Q - 1)\pi \leq w < (2Q+ 1)\pi $ for the topological sector with index  $Q$.

The R\'enyi entropies \eqref{LowTEEW} are depicted in FIG. \ref{fig:RenyiLowT2} for $n=3,2,1$ in the units of $2/ \sinh (\beta/2) $ for $l =1$. They show a linearly growing tendency (due to $ 2Q R_l \sin (R_l)$ term) with oscillating contributions. We note the R\'enyi entropy with larger number of replica copies start with smaller values and grow faster as $w$ increases, which is evident in the inset.  

In closing, we note that our analyses reveal that both the chemical potential and Wilson loop play significant roles, especially when $\mu=N/2$ and $w= (2Q\-1) \pi/n$, respectively, for different integers $N$ and $Q$. Careful examinations reveal that these two do not interfere with each other. Thus, chemical potential can still be used to probe the ground state energy levels of compact fermions even in the presence of Wilson loop, which accommodate first order phase transitions in this article, and vice versa. \\  \vspace{-0.05in}

\noindent {\textbf{IV. Measurements:}} \\
In \cite{Goldstein:2017bua}, the authors have performed numerical computations of the entropies with different normalizations for 2 dimensional fermions. We note that threading unalike Aharonov-Bohm flux $\alpha$ (same as our $w$) for different replica sheets as in FIG. 1 of \cite{Goldstein:2017bua} is distinct from our Wilson loop configurations. Here, we apply the same Wilson loop for all replica fermions that is restricted by our formalism. While our formulation is different from theirs, it will be interesting to explicitly check whether various entropies with same setup agree with each other or not. Moreover, the authors suggested possible experiments by slightly modifying the established experiment \cite{Experiment}. It will be of great interest to verify our results along with theirs in \cite{Goldstein:2017bua} by performing the suggested experiments. \\   

{\it Acknowledgements:} 
We thank John Cardy, Paul de Lange, Mitsutoshi Fujita, Elias Kiritsis, and Alfred Shapere  
for helpful discussions. We are especially thankful to Sumit Das for numerous discussions and valuable comments. We also thank to Moshe Goldstein for pointing out the possibility of experimental measurement of the charged R\'enyi entropy.

\onecolumngrid \vspace{0.2in}

\section*{\Large Supplementary material} 

\subsection{Partition \& Correlation functions} \label{sec:PartitionFunction}

To compute the partition function, we consider a torus with the modular parameter $\tau = \tau_1 + i \tau_2$. Thus the space of coordinates $\zeta = \frac{1}{2\pi} (s + i t)$ is identified as $\zeta \equiv \zeta + 1 \equiv \zeta + \tau$. Now $s$ is a spatial coordinate with a circumference $2\pi L$ (with $L=1$ in this Supplementary material), while $t$ is the Euclidean time with periodicity $2\pi \tau_2 = \beta = 1/T$. 

There are equivalent descriptions\cite{DiFrancesco:1997nk}\cite{Hori:2003ic} between the Dirac fermion $\psi$ in the presence of the twisted boundary conditions 
\begin{align}\label{TwistedeBC}\tag{S1}
	\psi (t,s) &= e^{-2\pi i \tilde a} \psi (t,s+2\pi) 
	= e^{-2\pi i \tilde b} \psi (t+2\pi \tau_2, s+2\pi \tau_1)  \;. 
\end{align}
and the fermions with flat gauge connection $\tilde A_\mu$ on a torus
\begin{align}\label{FlatConnection}\tag{S2}
	\tilde A = \tilde A_\mu dx^\mu = \tilde a ds + \frac{\tilde b -\tilde a\tau_1}{\tau_2} dt \;. 
\end{align}
Thus, in this equivalent description, one can identify the chemical potential and current source 
\begin{align} \tag{S3}
	\tilde a = \tilde J \;,  \qquad \tilde b = \tau_1 \tilde J + i \tau_2 \tilde \mu \;.
\end{align}
This can be used to construct the partition and correlation functions with background gauge fields in terms of twist boundary conditions \cite{Kim:2017ghc}. 

Partition function is a trace of the Hilbert space constructed with the twisted periodic boundary condition for $s \sim s + 2\pi$ along with the Euclidean time evolution $t \to t +2\pi \tau_2$ represented by the operator $ e^{-2\pi \tau_2 H}$, where $H$ is Hamiltonian. The latter also induces the space translation $s \to s - 2\pi \tau_1$ represented by the operator $ e^{-2\pi i \tau_1 P}$ (with momentum operator $P $) together with a phase rotation due to the presence of fermion $ e^{-2\pi i (\tilde b-1/2) F_A}$, where $F_A $ is a fermion number. The partition function with the twisted parameters $\tilde a$ and $\tilde b$ is 
\begin{align} \tag{S4}
	Z_{[\tilde a, \tilde b]} = Tr \big[ e^{-2\pi i (\tilde b-1/2) F_A} e^{-2\pi i \tau_1 P} e^{-2\pi \tau_2 H} \big] = \left|\eta(\tau) \right|^{-2} \big|\vartheta \genfrac[]{0pt}{1}{1/2- \tilde a}{\tilde b-1/2 } (0,\tau)\big|^2
	\;,
\end{align}   
where Dedekind function $\eta(\tau) = q^{\frac{1}{24}} \prod_{n=1}^\infty (1 - q^n)$, Jacobi function $\vartheta \genfrac[]{0pt}{1}{\alpha}{\beta} (z|\tau) = \sum_{n \in \mathbb Z} q^{\frac{(n + \alpha)^2}{2}} e^{2\pi i (z+\beta)(n + \alpha)}$, and $ q= e^{2\pi i \tau}$. 
The following notations are also used in the literature: $\vartheta_3(z|\tau) = \vartheta \left[\substack{0 \\ 0} \right] (z|\tau)$, $\vartheta_2(z|\tau) = \vartheta \left[\substack{1/2 \\ 0} \right] (z|\tau)$, $\vartheta_4(z|\tau) = \vartheta \left[\substack{0 \\ 1/2} \right] (z|\tau)$, and $\vartheta_1(z|\tau) = \vartheta \left[\substack{1/2 \\ 1/2} \right] (z|\tau) $. $\vartheta_2(z|\tau)$ is related to the periodic spatial circle and anti-periodic time circle, while $\vartheta_3(z|\tau)$ to both the anti-periodic spatial and time circles. $\vartheta_2$ and $\vartheta_3$ are physical fermions described by the anti-periodic temporal circle.

In the presence of current source $J$ and chemical potential $\mu$, the partition function is 
\begin{align} \tag{S5}
	Z_{[\tilde a, \tilde b]}^{\mu, J} = Tr \left[e^{2\pi i (\tau_1 J + i \tau_2 \mu + \tilde b-\frac{1}{2}) F_A} e^{\-2\pi i \tau_1 P} e^{\-2\pi \tau_2 H} \right] 
	=\left|\eta(\tau) \right|^{-2} \big|\vartheta \genfrac[]{0pt}{1}{1/2-\tilde a- J}{\tilde b-1/2 } (\tau_1 J +i\tau_2 \mu|\tau)\big|^2
	\;.
\end{align}   
Note that the current source has two distinct contributions. One of them is through the modification of the Hilbert space. 
The periodicity of temporal direction and the corresponding thermal boundary condition are not modified. With this we can construct the two point correlation functions in the presence of chemical potential $\mu$ and current source $J$ following \cite{DiFrancesco:1997nk}. The correlation functions have been explicitly written in \cite{Kim:2017ghc}. 

Now we construct the partition functions for $n$-replica fermions in the presence of the Wilson loop. Without the Wilson loop, we impose the identifications when the $m$-th fermion $\tilde \psi_m$ ($m=1,2,\cdots, n$) crosses the branch cut connecting two points $u$ and $v$ in 2 dimensional Riemann space,  
\begin{align} \tag{S6}
	\tilde \psi_m(e^{2\pi i} (x-u)) = \tilde \psi_{m+1} (x-u) \;, \qquad \tilde \psi_m(e^{2\pi i} (x-v)) = \tilde \psi_{m-1} (x-v) \;,
\end{align}	  
where $x$ is a complexified coordinate. 
These boundary conditions can be diagonalized by defining $n$ new fields  
\begin{align} \tag{S12}
	\psi_k = \frac{1}{n} \sum_{m=1}^n e^{2\pi i m k } \tilde \psi_m  \;.
\end{align}	 
For the new field, the boundary condition becomes 
\begin{align} \tag{S7}
	\psi_k(e^{2\pi i} (x-u)) = e^{2\pi i k/n} \psi_{k} (x-u) \;, \qquad \psi_k(e^{2\pi i} (x-v)) = e^{-2\pi i k/n} \psi_{k} (x-v) \;,
\end{align}	 
where $ k=-(n-1)/2, -(n-3)/2, \cdots, (n-1)/2$. The phase shift $e^{2\pi i k/n}$ is generated by the standard twist operator $\sigma_{k/n}$ with a conformal dimension $ \frac{1}{2} \frac{k^2}{n^2}$. The full twist operator is $\sigma_n = \Pi  \sigma_{k/n}$. 

In \cite{Belin:2013uta}, the authors notice that the boundary condition can be further generalized to include a global phase rotation $\tilde \psi \to e^{i w} \tilde \psi$. This phase can be added to the boundary condition for $\tilde \psi$ as 
\begin{align} \tag{S8}
	\tilde \psi_m(e^{2\pi i} (x-u)) = e^{i w} \tilde \psi_{m+1} (x-u) \;, \qquad \tilde \psi_m(e^{2\pi i} (x-v)) = e^{-i w} \tilde \psi_{m-1} (x-v) \;.
\end{align}	  
This additional phase is added uniformly, the diagonal fields has the same effect. 
\begin{align}\label{GeneralTWB} \tag{S15}
	\psi_k(e^{2\pi i} (x-u)) = e^{2\pi i k/n+iw} \psi_{k} (x-u) \;, \qquad \psi_k(e^{2\pi i} (x-v)) = e^{-2\pi i k/n-iw} \psi_{k} (x-v) \;.
\end{align}	 

These phase shifts can be handled by the generalized twist operators $\sigma_{w,k}$ with an electric parameter $w$ and a magnetic parameter $k$. Their name for the electromagnetic operators come from \cite{DiFrancesco:1997nk} along with the detailed expressions for the correlation functions. The expression for the two point functions of the electromagnetic operators $ \sigma_{w,k} (u)$ and $\sigma_{-w,-k} (v)$ with an electric charge $\frac{w}{2\pi} + l_k$ and a magnetic charge $\frac{k}{n}$, in the presence of the current source $J$ and the chemical potential $\mu$ as well as the twisted boundary conditions $\tilde a, \tilde b$, are given by  
\begin{align} \label{TopologicalCorrelatorTwistedOperators} \tag{S9}
	\langle \sigma_{w,k} (u) \sigma_{-w,-k} (v) \rangle_{\tilde a,\tilde b,J,\mu} = \Big| \frac{2\pi \eta (\tau)^3 }{\vartheta [\substack{1/2 \\ 1/2 }](\frac{u-v}{2\pi L}|\tau)} \Big|
	^{2\alpha_{w,k}^2}~ 
	\Big| \frac{\vartheta [\substack{1/2-\tilde a-J \\ \tilde b-1/2 }](\frac{u-v}{2\pi L} \alpha_{w,k}+ \tau_1 J + i \tau_2 \mu|\tau)}{\vartheta [\substack{1/2-\tilde a-J \\ \tilde b-1/2 }](\tau_1 J + i \tau_2 \mu|\tau)} \Big|^2 \;. 
\end{align}
where $\alpha_{w,k} = \frac{k}{n} + \frac{w}{2\pi} + l_k$. The magnetic parameter $\frac{k}{n}$ comes into a play when we use the replica trick for $n$ copies of fermions in a single Riemann surface instead of a fermion on $n$ copies of Riemann surfaces \cite{Casini:2009sr}. The corresponding conformal dimension is given by 
\begin{align} \label{ConformalDimension} \tag{S10}
	\Delta_{w,k}= \text{conformal dimension} = \frac{1}{2}  \alpha_{w,k}^2= \frac{1}{2} \left( \frac{k}{n} + \frac{w}{2\pi} + l_k \right)^2  \;.
\end{align}	 
The constant $l_k$ stems from the fact that the boundary condition \eqref{GeneralTWB} has an intrinsic ambiguity. This constant is used to minimize the conformal dimension of the twist operator such that $ -\frac{1}{2} \leq \alpha_{w,k} \leq \frac{1}{2}$. This is closely related to the topological transitions discussed in the main body. It is interesting to mention that the dependence of the background gauge fields in entanglement entropy can be systematically introduced by using the twisted boundary conditions represented by the parameters $\tilde a$ and $\tilde b$ \cite{Kim:2017ghc}\cite{Kim:2018xla} .

\subsection{Example Computations for anti-periodic fermions in the zero temperature limit} \label{sec:AntiPeriodicZeroEE}

We derive the equation (12) in the main body. Entanglement entropy for the anti-periodic fermions is given by 
\begin{align}\tag{S11}
	S_{n= 1}^\mu &= \lim_{n\to 1} \Big(  \frac{1}{1- n} \bigg[ \sum_{k=-\frac{n-1}{2}}^{\frac{n-1}{2}} \log \Big|\frac{\vartheta_3 (\frac{k}{n} \frac{\ell_t}{2\pi L}+\frac{i\beta\mu}{2\pi} |\frac{i\beta}{2\pi})}{\vartheta_3 (\frac{i\beta\mu}{2\pi} |\frac{i\beta}{2\pi})} \Big|^2 \bigg] \Big) \;, 
\end{align} 
where $\beta = 2\pi \tau_2$ and $\ell_t = u-v $. 
Using the product representation 
$$
\vartheta_3 (z|\tau) = \prod_{m=1}^\infty (1 - q^m)(1 + y q^{m-1/2})(1 + y^{-1} q^{m-1/2}) \;,
$$
with $y_1 = e^{- \beta\mu + i \frac{k}{n} \frac{\ell_t}{L}}, ~y_2 = e^{- \beta\mu}, ~ q=e^{-\beta}$. \\

We compute the R\'enyi entropy at low temperature limit, $\beta \to \infty$, 
\begin{align}\label{app2}\tag{S12}
	\begin{split}	
		S_n^\mu &= \frac{1}{1- n} \bigg[ \sum_{k=-\frac{n-1}{2}}^{\frac{n-1}{2}} \log \Big|\prod_{m=1}^{\infty} \frac{(1 - q^m)(1 + y_1 q^{m-1/2})(1 + y_1^{-1} q^{m-1/2})}{(1 - q^m)(1 + y_2 q^{m-1/2})(1 + y_2^{-1} q^{m-1/2})} \Big|^2 \bigg]  \\
		&= \frac{1}{1- n} \bigg[ \sum_{m=1}^{\infty} \sum_{k=-\frac{n-1}{2}}^{\frac{n-1}{2}} \sum_{l=1}^{\infty} \frac{(-1)^{l-1}}{l} \left( [y_1^l +y_1^{l*} - y_2^l - y_2^{l*}] q^{l(m-1/2)} \right. \\
		&\left.\hspace{2.2in} + [y_1^{-l} + y_1^{-l*} - y_2^{-l}  - y_2^{-l*} ]q^{l(m-1/2)}  \right) \bigg]    \\
		&= \frac{2}{1- n} \bigg[ \sum_{l,m=1}^{\infty} \sum_{k=-\frac{n-1}{2}}^{\frac{n-1}{2}}  \frac{(-1)^{l-1}}{l} 
		\left( e^{-l\beta\mu} + e^{l\beta\mu}\right) e^{- l\beta(m-1/2)} \Big( \cos \big(\frac{k}{n} \frac{l \ell_t}{L}\big) -1 \Big)  \bigg]   \\
		&= \frac{2}{1- n} \bigg[ \sum_{l,m=1}^{\infty} \frac{(-1)^{l-1}}{l} 
		\left( e^{-l\beta\mu} + e^{l\beta\mu}\right) e^{- l\beta(m-1/2)} \Big(-n +  \sin \big( \frac{l \ell_t}{2 L}\big) \csc \big(\frac{1}{n} \frac{l \ell_t}{2 L}\big) \Big)  \bigg]  \;.
	\end{split}	
\end{align}
In the second line, the product associated with the index $m$ turns into a sum outside the $\log$ and the index $l $ runs for the expansion $\log (1 + y q^{m-1/2}) = \sum_l (-1)^{l-1} y^l q^{l(m-1/2)}/l$ for small $q$. \\

Entanglement entropy can be computed by taking the $n\to 1$ limit, which is straightforward. 
\begin{align} \tag{S13}
	S_{n= 1}^\mu &=2\sum_{l=1}^{\infty} \frac{(-1)^{l-1}}{l} 
	\frac{\cosh\left(l\beta\mu\right)}{\sinh\left(l\beta/2\right)} \left[1- \frac{l \ell_t}{2 L} \cot \left(\frac{l \ell_t}{2L} \right) \right]  
	\;. \nonumber 
\end{align} 

\end{document}